\documentclass{iau}
\usepackage{}
\usepackage{amssymb}
\usepackage{graphicx}

\title[sps13.~RS Cvn eclipsing binary DV Psc]
{Photometric study of the short-period \\RS CVn eclipsing binary DV Psc}

\author [Qing-feng Pi, Li-yun Zhang, Zhongmu Li et al.,]
{Qing-feng Pi$^1$ Li-yun Zhang$^1$ Zhongmu Li$^2$ Hong-yan Sang$^3$ \and Zhongzhong Zhu$^1$}

\affiliation{$^1$College of Science/Department of Physics, Guizhou
University, Guiyang 550025, P.R. China \\e-mail: Liy\_zhang@hotmail.com or piqingfeng@126.com
\\ $^2$Astronomy and History of Science and Technology, Dali University, Dali 671003, P.R. China
\\ $^3$School of Mathematics Science, Liaocheng University, Liaocheng, 252059, China}

\pubyear{2012}
\volume{xxx}  %% insert here IAU Symposium No.
\pagerange{119--126}
\jname{IAU Sp. Ses. 13}
\editors{David Soderblom, eds.}
\begin{document}

\maketitle

\begin{abstract}
We present our new photometry of DV Psc obtained in 2010 and 2011, and new spectroscopic observation on
Feb. 14, 2012. During our observations, three flare-like events might be detected firstly in
one period on DV Psc. The flare rate of DV Psc is about 0.017 flares per hour. Using Wilson-Devinney program,
we derived the preliminary starspot parameters.
Moreover, the magnetic cycle is 9.26($\pm0.78$) year analyzed by variabilities of Max.I - Max.II.

%% add here a maximum of 10 keywords, to be taken form the file <Keywords.txt>
\end{abstract}

\firstsection

\section{Introduction}
DV Psc is a RS CVn-type eclipsing binary with high-level surface activity, which is characterized
by the light-cuvre asymmetries (Robb et al., 1999; Vanko et al., 2007; Zhang \& Zhang, 2007; Zhang
et al., 2010; Parimucha et al.,2010) and $\mbox{Ca~{\sc ii}}$ H \& K emission (Beer 1994). Therefore,
it is a very intriguing object for studying stellar magnetic activity.

\section{Light-curve and spectral analysis}
Our new CCD photometric observations of DV Psc (Fig. 1) were made in four observing runs:
2010 Nov. 19, 20, and 2011 Oct. 27, Nov. 12 and 13, Dec. 08 and 10 with 85cm telescope (Zhou et al., 2009) at Xinglong
station of the National Astronomical Observatories of China ($\emph{NAOC}$). The spectroscopic observations of DV Psc were
obtained with the 2.16m telescope at Xinglong station on 2012 Feb. 14 (Fig. 2). The OMR spectrograph centered at about 4280
${\AA}$ with a reciprocal dispersion of 1.03 ${\AA}$ (Fang et al., 2010).\\
\indent Multi-color light-curve analyses were carried out using the Wilson-
Devinney program (Wilson \& Devinney 1971; etc). The spot models
are used to explain light-curve asymmetry based on the prior photometric solution (Zhang et al.
2010). The theoretical and observed light-curve are both displayed in Figure 1. The normalized spectra of DV Psc were analyzed
in the $\mbox{Ca~{\sc ii}}$ H \& K, H$_{\gamma}$ and H$_{\beta}$ lines with the spectral subtraction technique, which was
described in detail by Barden (1985) and Montes et al. (1995). \\

\begin{figure}
% \vspace*{-2.0 cm}
\begin{center}
\includegraphics[width=1in]{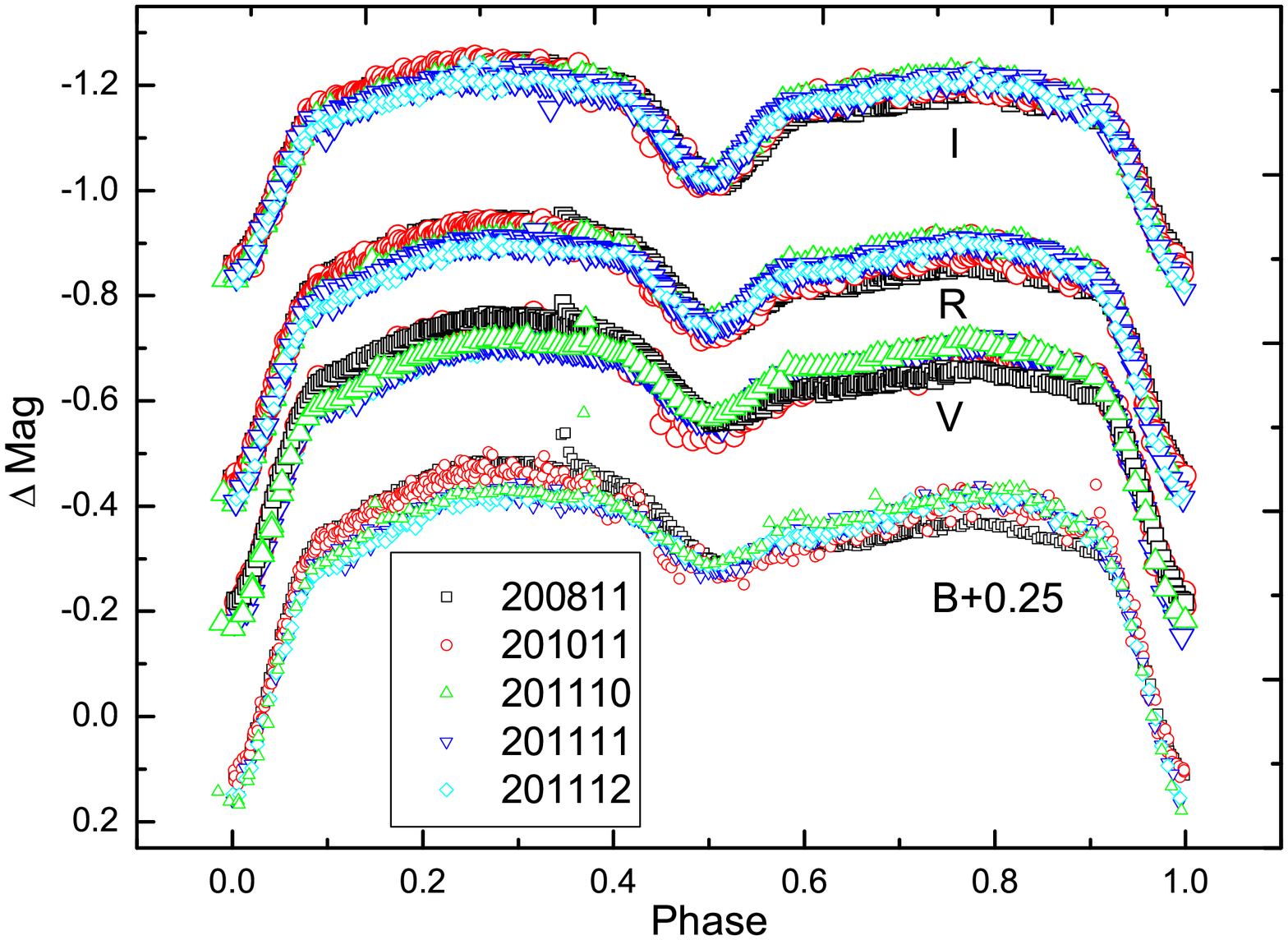}
\includegraphics[width=1in]{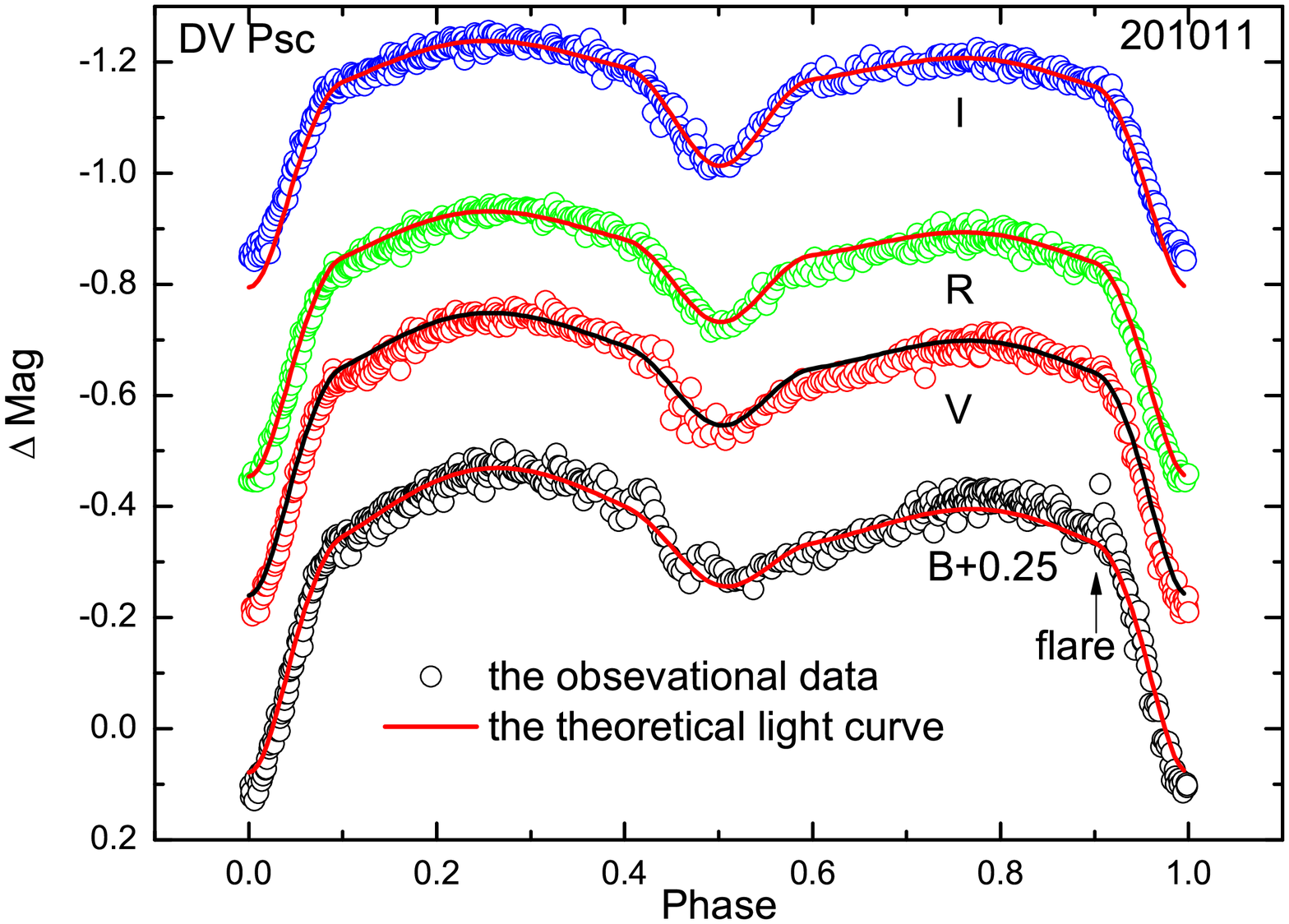}
\includegraphics[width=1in]{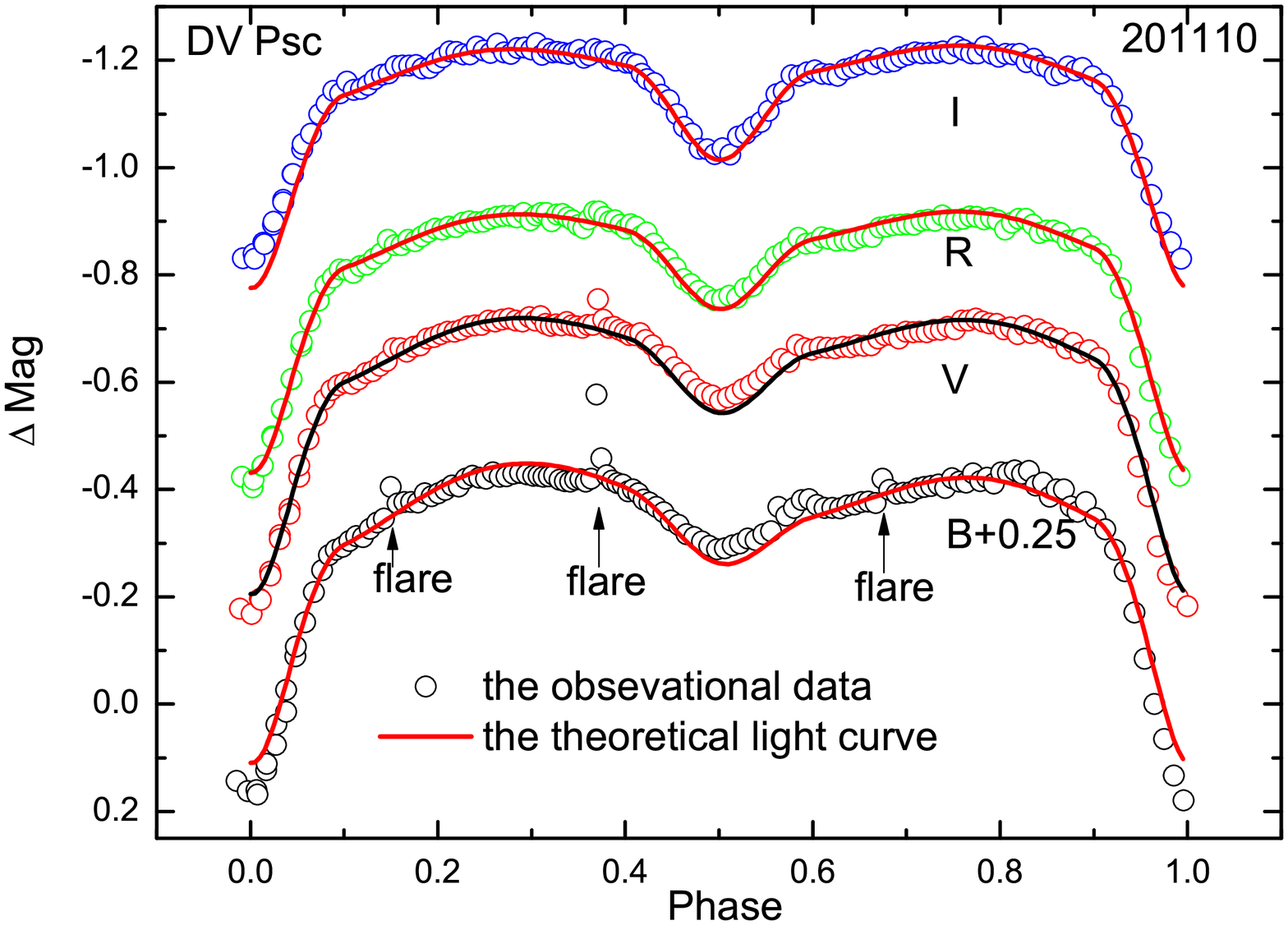}
\includegraphics[width=1in]{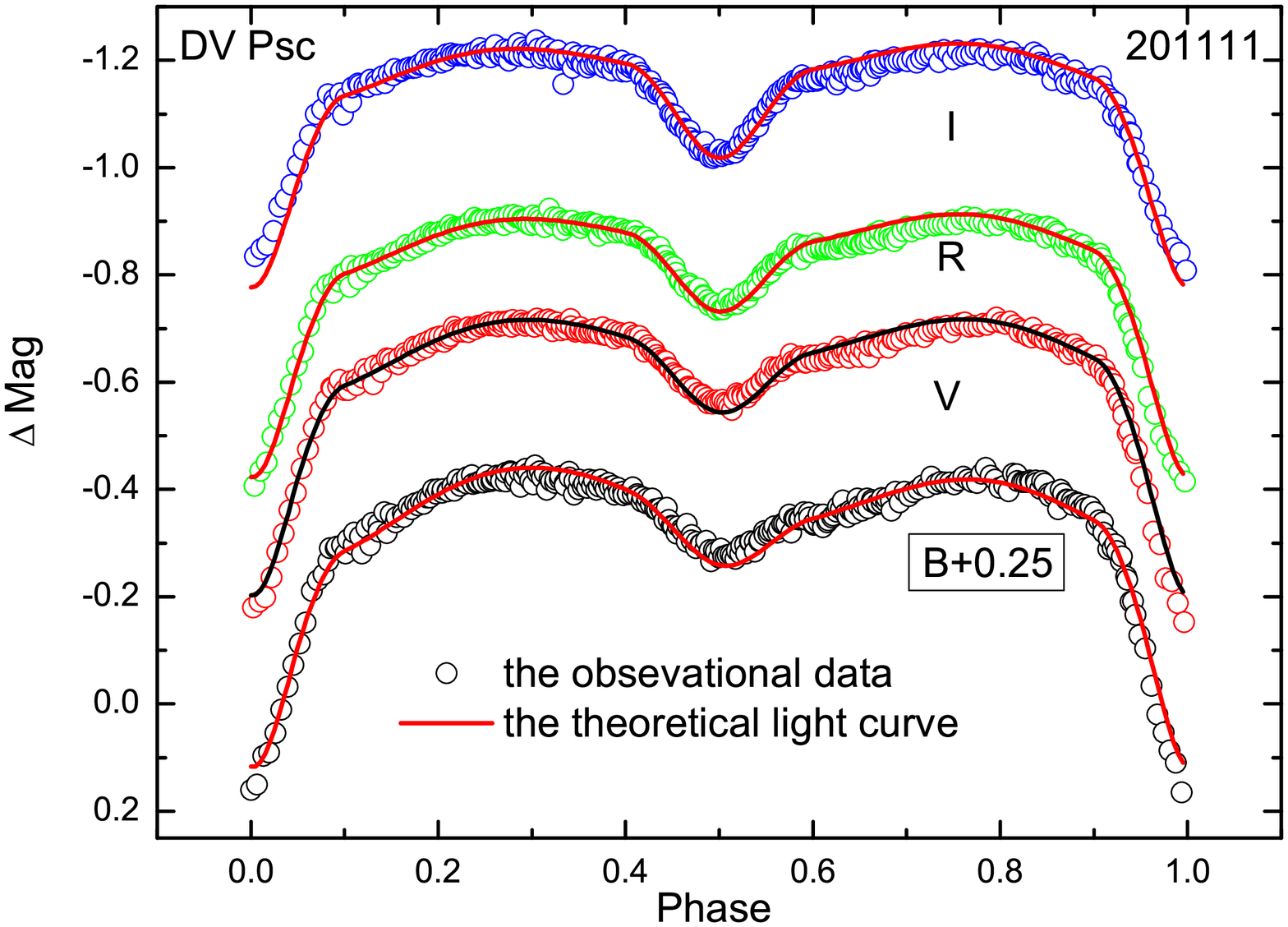}
\includegraphics[width=1in]{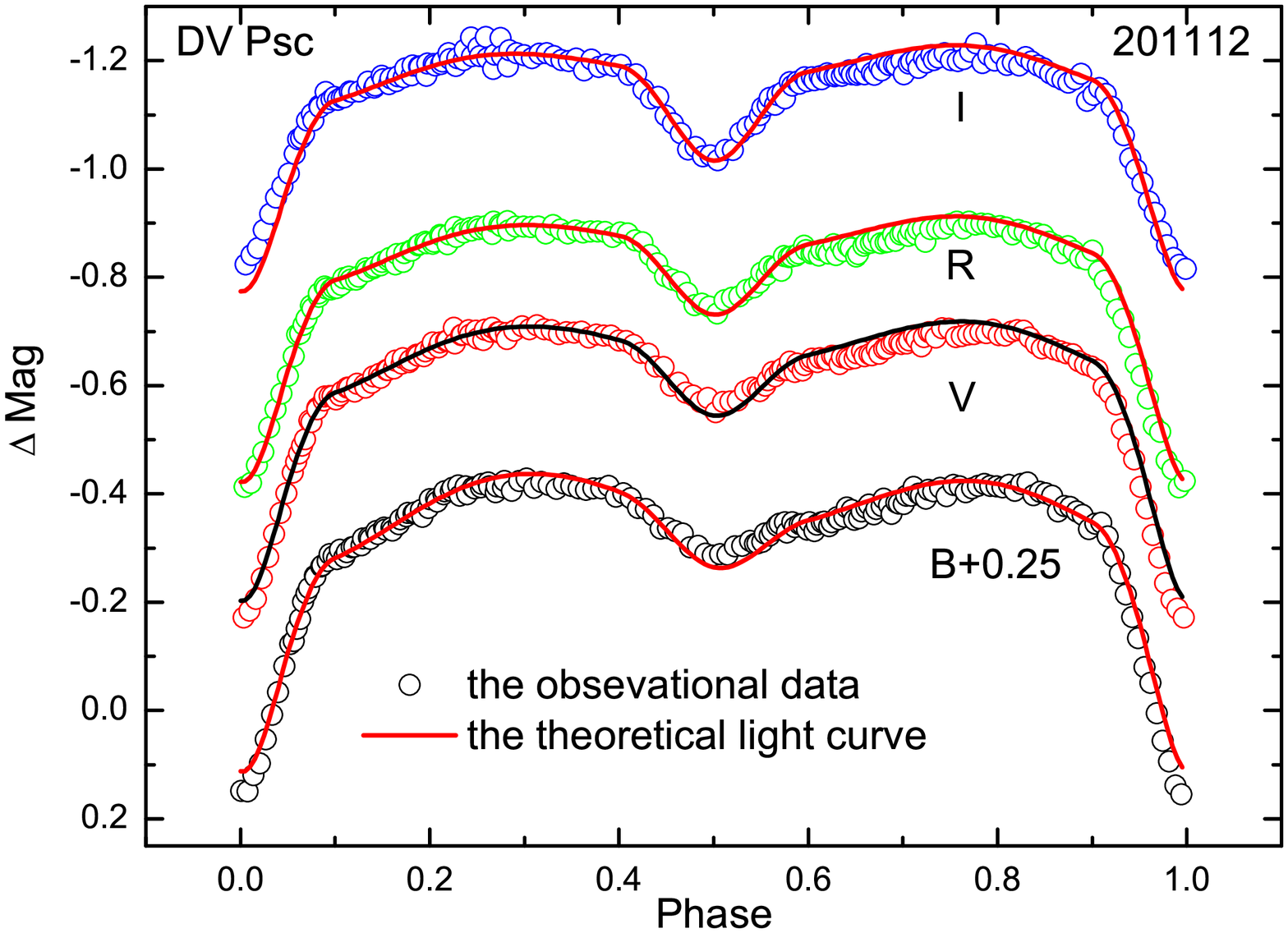}
% \vspace*{-1.0 cm}
\caption{The
light-curves with their best fits for DV Psc in \emph{BVRI} bands.}
\label{fig}
\end{center}
\end{figure}
\begin{figure}
% \vspace*{-2.0 cm}
\begin{center}
\includegraphics[width=1.0in]{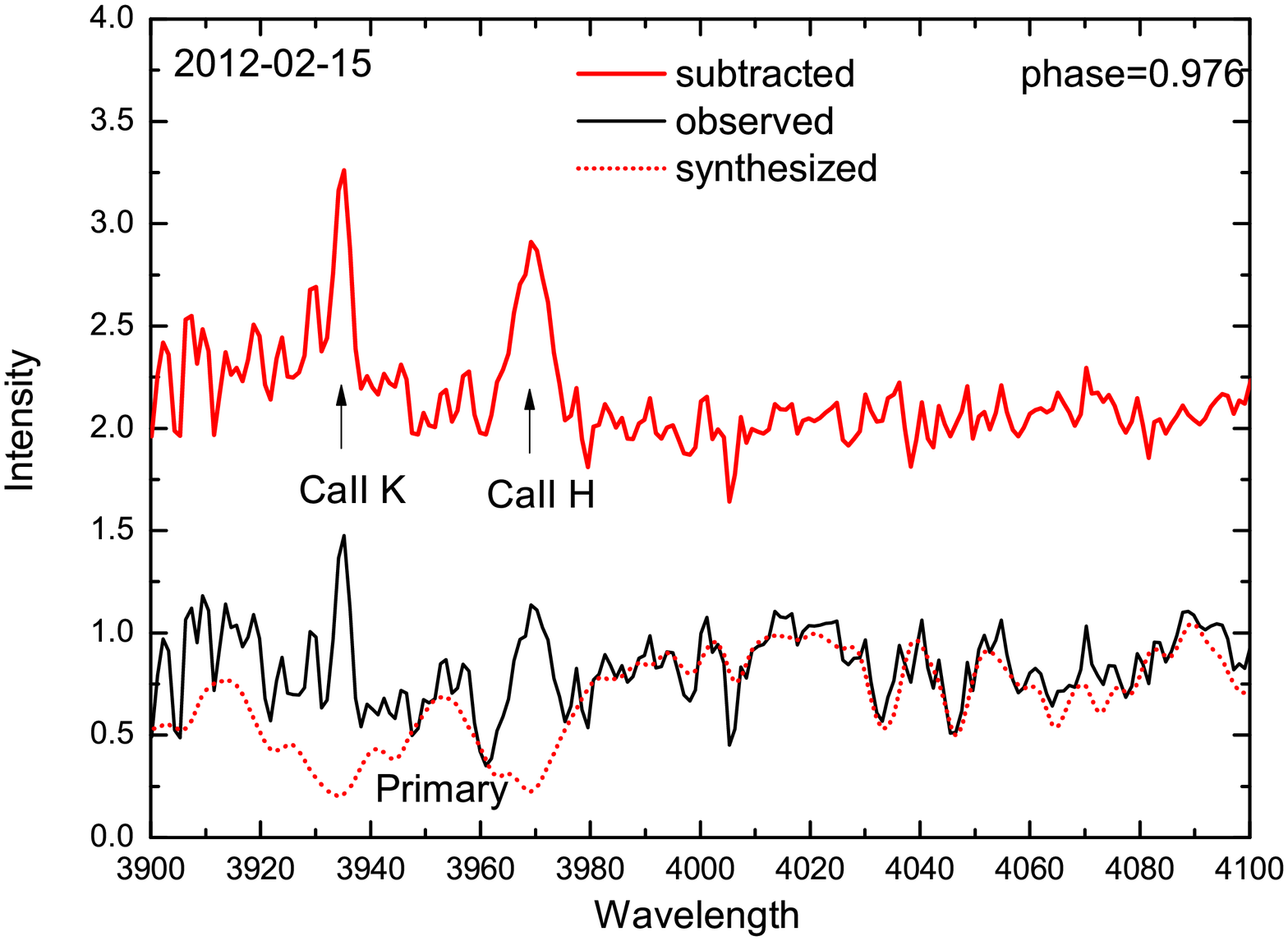}
\includegraphics[width=1.0in]{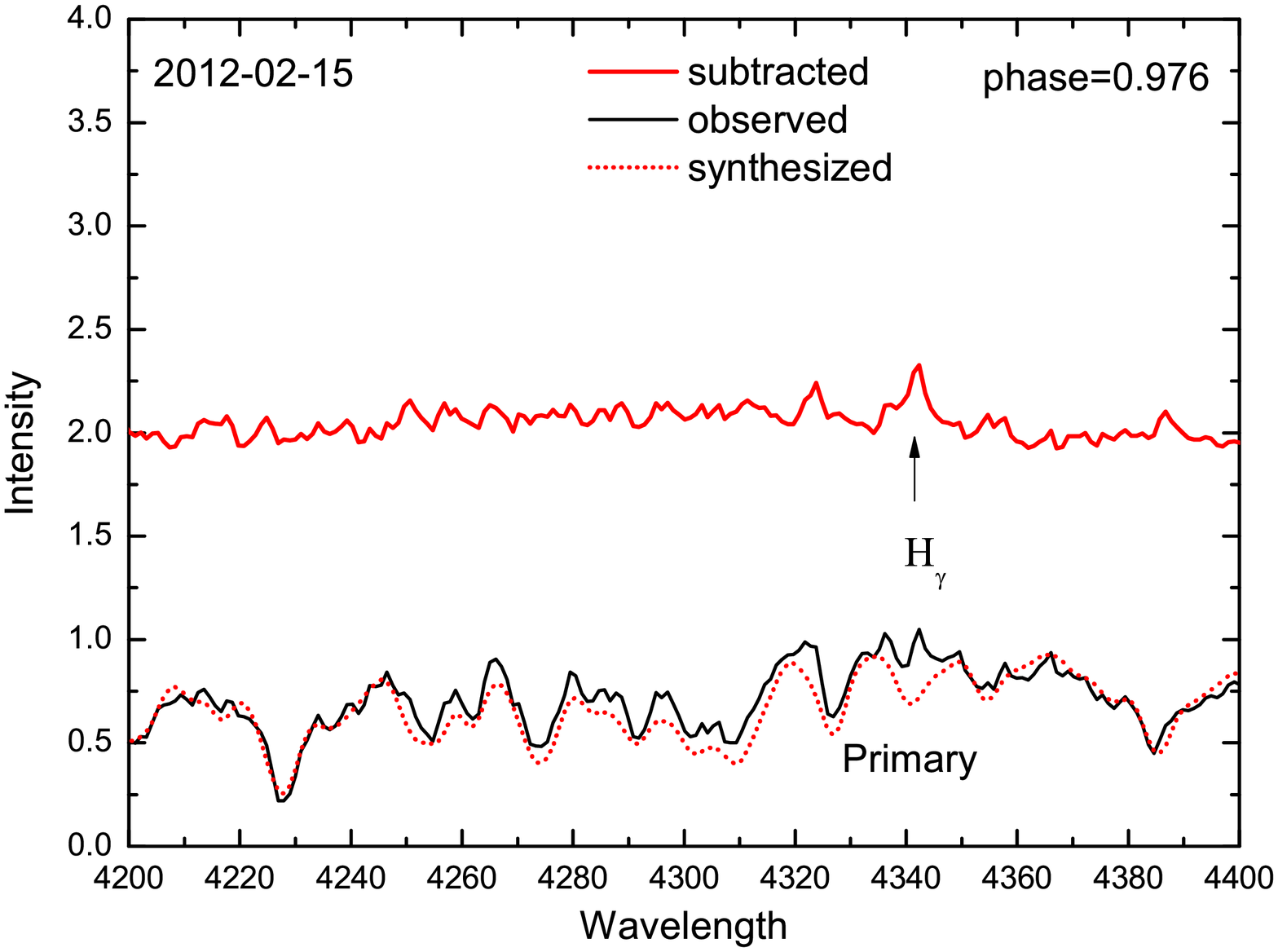}
\includegraphics[width=1.0in]{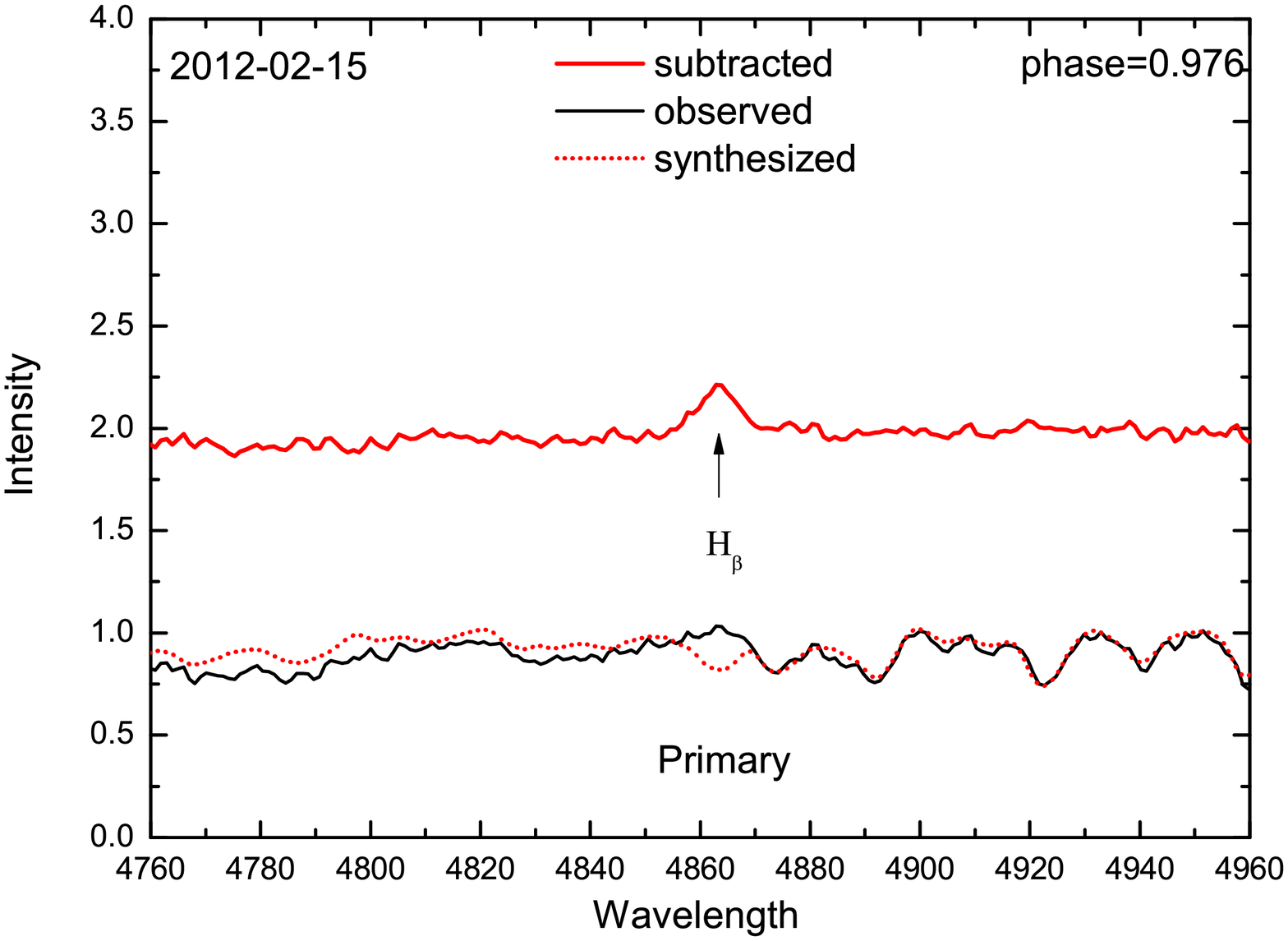}
% \vspace*{-1.0 cm}
\caption{The observed, synthesized, and subtracted
spectra for the $\mbox{Ca~{\sc ii}}$ H\&K, H$_{\beta}$ and H$_{\gamma}$ lines. The dotted
lines represent the synthesized spectra and the upper are
the subtracted ones.}
\label{fig}
\end{center}
\end{figure}

\begin{figure}
% \vspace*{-2.0 cm}
\begin{center}
\includegraphics[width=1in]{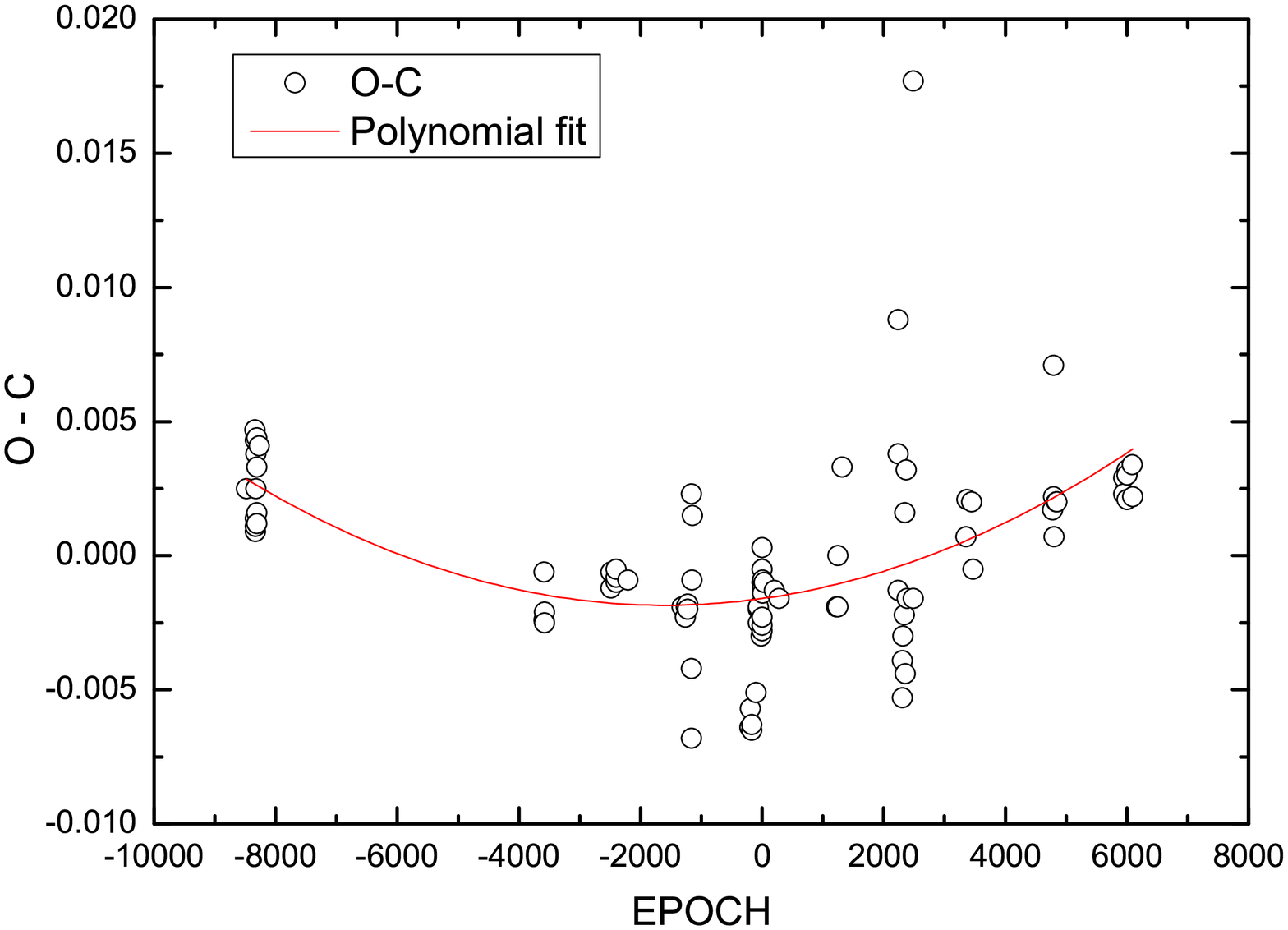}
\includegraphics[width=1in]{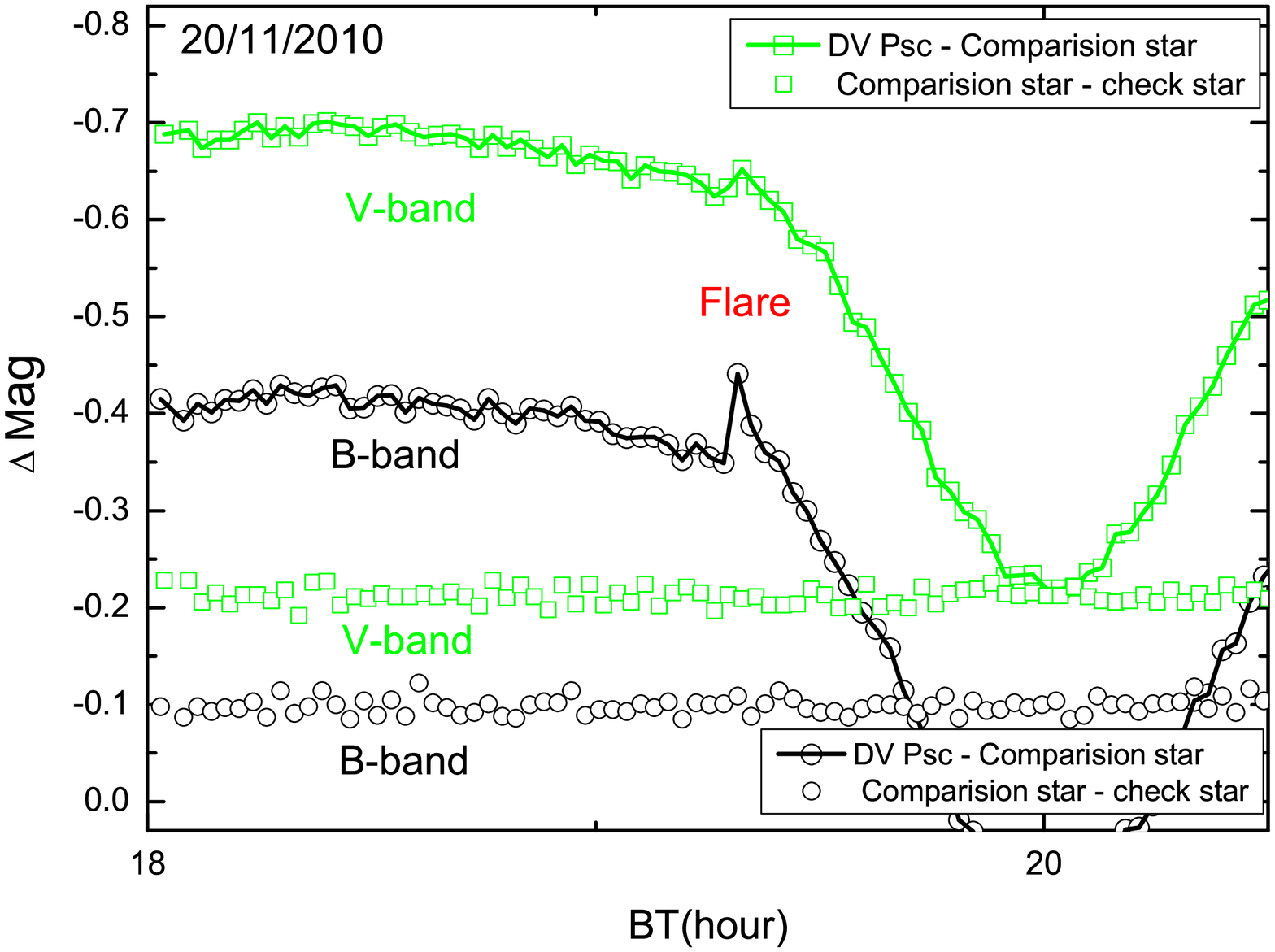}
\includegraphics[width=1in]{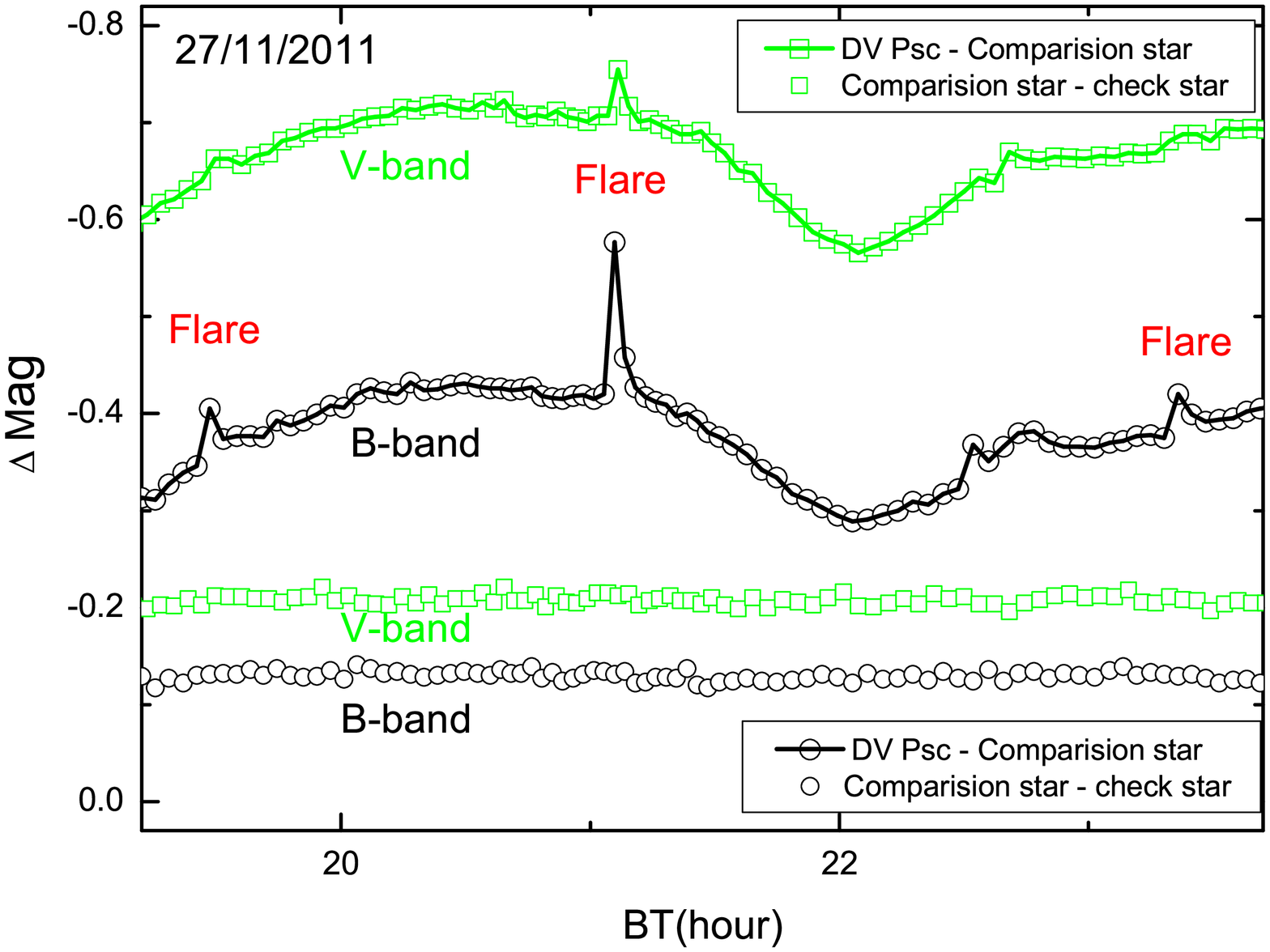}
\includegraphics[width=1in]{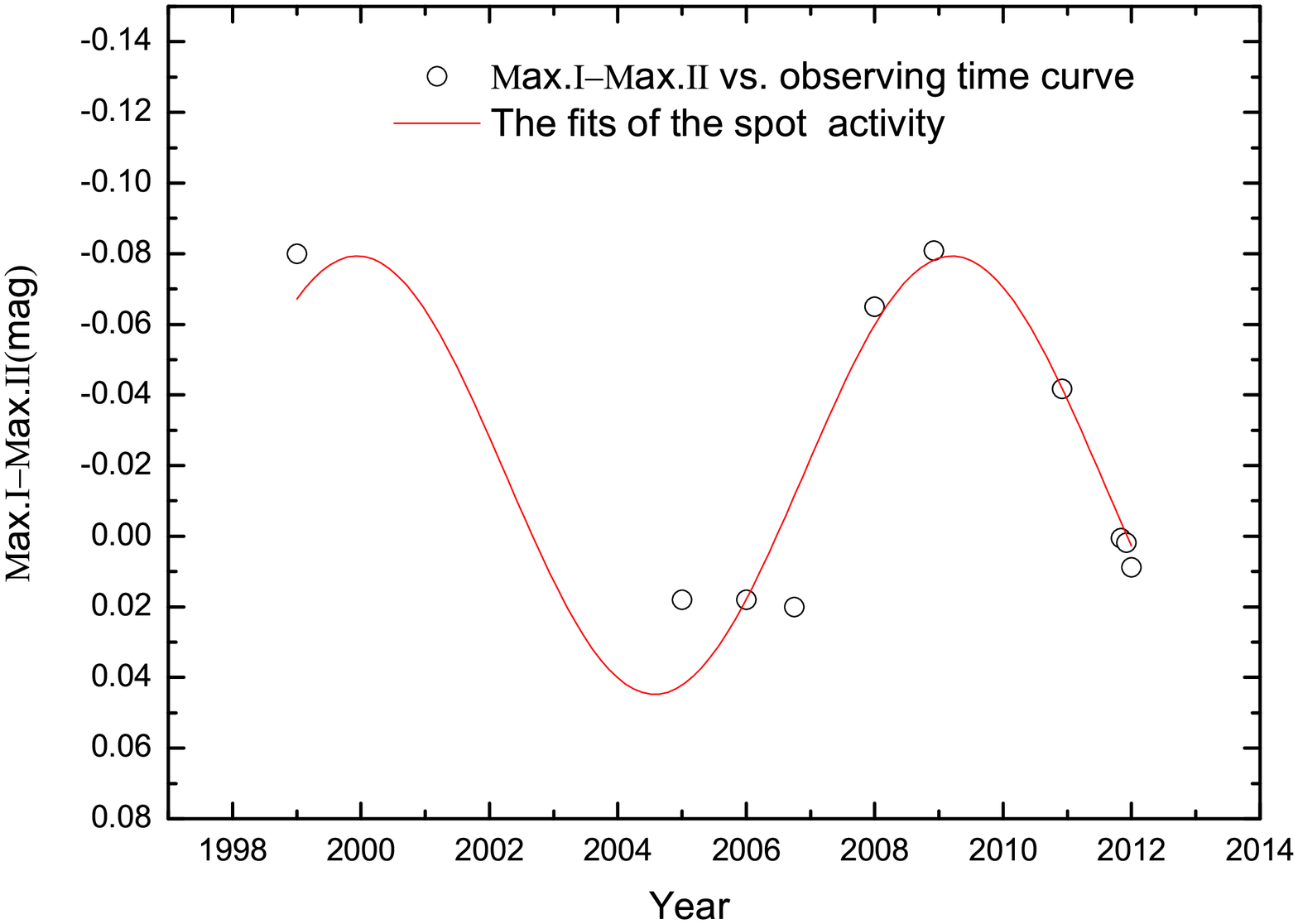}
\includegraphics[width=1in]{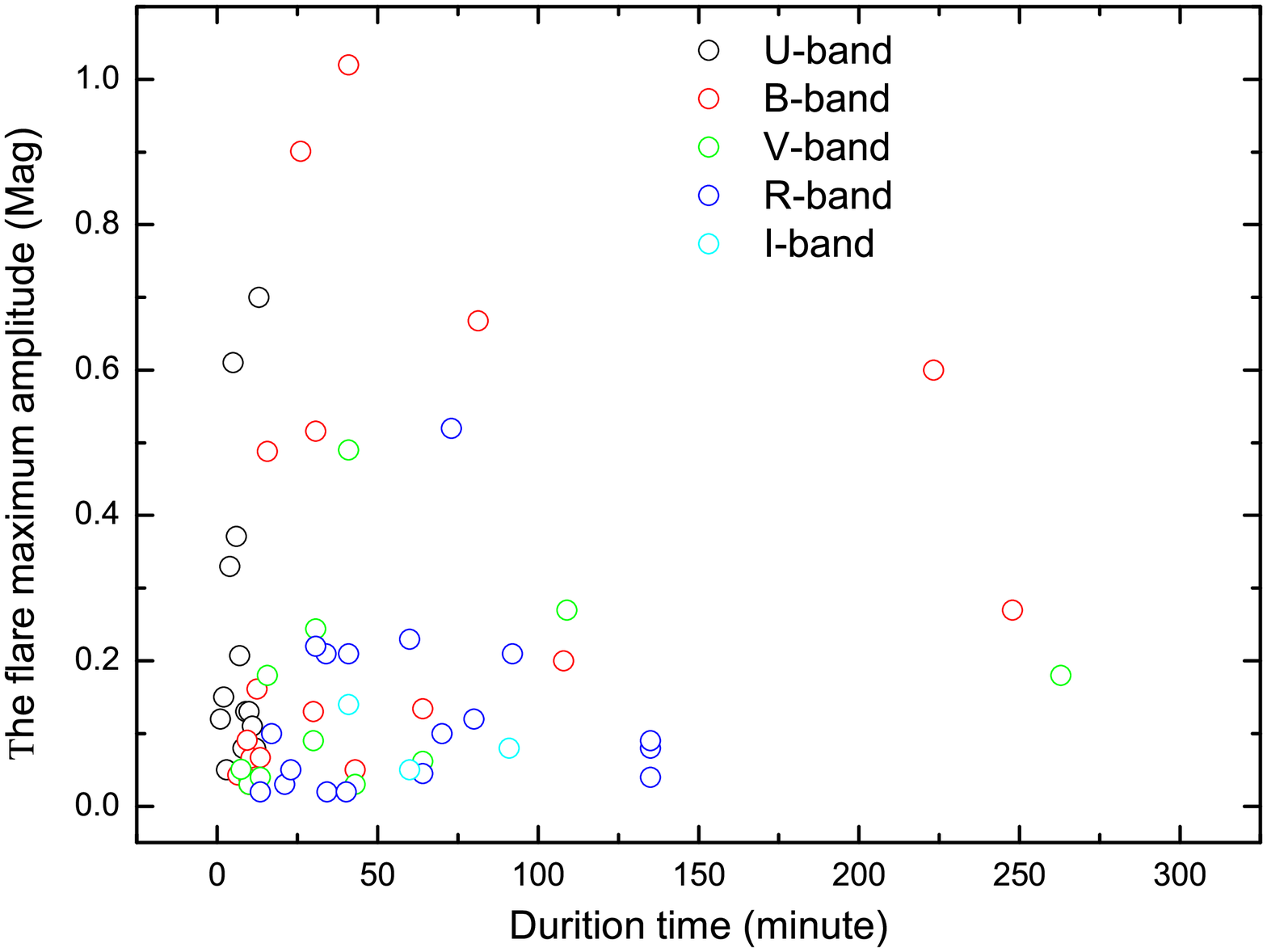}
% \vspace*{-1.0 cm}
\caption{(O - C) diagram, the flares in \emph{B} and \emph{V} band, and magnetic cycle for DV Psc.
The relation of the flare amplitude and duration of late-type stars.}
\label{fig}
\end{center}
\end{figure}

\section{Conclusion}

The results can be summarized as follows: 1. An updated linear ephemeris formula Min.I~=~JD(Hel.)2454026.1424(2) + 0.30853609(8)E was obtained. Fitting all available light minimum times with a
polynomial function showed that the orbital period of DV Psc increased (Fig. 3). 2. Our photometric
and spectral results demonstrate that DV Psc is very active. New four flare events of DV Psc were found
and the flare rate is about 0.017 flares per hour (Fig. 3). The three flare-like events might be
detected firstly in one period. The relation of the flare maximum amplitude and the flare duration
of late-type stars (Kozhevnikova et al., 2006; Vida et al., 2009; Qian et al., 2012; Zhang et al. 2012, etc) are displayed
in Figure 3. 4. The magnetic active cycle may be 9.26($\pm$0.78) year, which was analyzed by the
variabilities of Max.I - Max.II (Fig. 3). We will monitor later.

\begin{acknowledgements}

We are very grateful to Dr. Montes D., Gu S. H., Han J. L., Zhou A. Y., Zhou X., Jiang X. J., Zhao Y. H., and Fang X. S. The
work is supported by the \emph{NSFC} under grant No. 10978010, 11263001, 11203005 and 10373023.
This work is partially Supported by the Open Project Program of the Key Laboratory of Optical
Astronomy, \emph{NAOC, CAS}.

\end{acknowledgements}

\end{document}